\documentstyle[epsf]{article}
\begin{document}
\pagenumbering{arabic}
{\noindent \Large \bf Microcanonical Transfer Matrix 
and Yang-Lee Zeros of the $Q$-State Potts Model}
\vskip 4mm
{\noindent \it Richard J. Creswick and Seung-Yeon Kim} 
\vskip 4mm
{\noindent Department of Physics and Astronomy, 
University of South Carolina,\\ Columbia, SC 29208, USA}

\section{Introduction}

The $Q$-state Potts model in two dimensions is very fertile ground
for the investigation of phase transitions and critical phenomena.
For $Q=2$ and 3 there is a second order phase transition
between $Q$ ferromagnetic ordered states and a disordered state.
For $Q=4$ the transition is also second order, but the usual
critical behavior is modified by strong logarithmic corrections.
For $Q>4$ the transition is first order, with $Q=5$ exhibiting
{\it weak} first order behavior and a very large correlation
length at the critical point.

The Hamiltonian for the Potts model is
$${\cal H}=J\sum_{<i,j>}(1-\delta(q_i,q_j)),\eqno{(1)}$$
where $J$ is the coupling constant and $0\le q_i \le Q-1$
are the Potts spin variables. The energies (in units of $J$)
are evenly spaced and take on interger values in the range
$0\le E \le N_b$ where $N_b$ is the number of bonds on the lattice.
Here we will consider simple square lattices with periodic,
cylindrical, and self-dual boundary conditions.

In addition to the energy there are $Q$ order parameters
$$M_q=\sum_{k}\delta(q_k,q),\eqno{(2)}$$
which for the Ising model is simply related to the magnetization.
The possible values of the order parameter are also intergers,
$0\le M \le N_s$, where $N_s$ is the number of sites on the lattice.

If we denote the number of states with energy $E$ by $\Omega(E)$,
then the canonical partition function for the $Q$-state Potts model is
$$Z_Q(y)=\sum_{E}\Omega_Q(E)y^E,\eqno{(3)}$$
where $y=e^{-\beta J}$. From (3) it is clear that $Z$ is simply 
a polynomial in $y$, and the analytic structure of $Z$ is completely
determined by the zeros of this polynomial, as first discussed
by Lee and Yang\cite{yl,ly}.

If we wish to study the partition function in an external field
which couples to the order paramter, (2), then one needs to
enumerate the number of states with fixed energy $E$ and
fixed order parameter $M$, $\Omega(E,M)$. The partition function
is again a polynomial given by
$$Z_Q(y,x)=\sum_{E}\sum_{M}\Omega_Q(E,M)x^M y^E,\eqno{(4)}$$
where $x=e^{-\beta h}$, and $h$ is the external field.
As discussed in the second\cite{ly} of Lee and Yang's two
famous papers, the zeros of the partition function for the Ising
model in the complex-$x$ plane all lie on the unit circle.

For finite systems the analyticity of $Z$ in both $x$ and $y$ ensures 
that no zeros lie on the real axis. However, according to Lee
and Yang, in the thermodynamic limit the zeros of the partition
function in either the complex-$x$ or -$y$ planes approach arbitrarily 
close to the real axis at the critical point, leading to
nonanalytic behavior in the partition function.

If the zeros lie on a one-dimensional locus 
in the thermodynamic limit one can define the 
density of zeros (per site) $g(\theta)$\cite{ab,su} 
in terms of which the free energy per site is
$$f(y)=-\int g(\theta)log[y-y_0(\theta)]d\theta.\eqno{(5)}$$
In the critical region the singular part of the free energy
is a homogeneous function of the reduced temperature, $y-y_c$,
from which it follows\cite{ck} that $g(\theta)$ must also be
a homogeneous function for small $\theta$ of the form
$$g(\theta)=b^{(-d+y_T)}g(\theta b^{y_T}).\eqno{(6)}$$
This in turn implies that $g(\theta)$ vanishes as 
$\theta^\kappa$ as $\theta$ goes to zero where $\kappa=(d-y_T)/y_T$.
On the other hand, if the system has a first order transition,
$\kappa=0$ and the discontinuity in the first derivative of $f$,
the latent heat, is given by $L=2\pi g(0)$. Exactly parallel
arguments hold in the complex-$x$ plane if one replaces
the temperature exponent $y_T$ by the magnetic exponent $y_h$. 

In a recent paper Creswick\cite{cr} has shown how the numerical
transfer matrix of Binder\cite{bi} can be generalized to allow
the evaluation of the density of states $\Omega(E)$ and the 
restricted density of states, $\Omega(E,M)$ for the $Q$-state
Potts model. Similar calculations of $\Omega(E)$ have been carried
out by Bhanot\cite{bh} in both two and three dimensions for 
the $Q=2$ and $Q=3$ Potts models, and Pearson\cite{pe}
for the $4^3$ Ising model. Beale\cite{be} has used the exact 
solution for the partition function of the Ising model on finite
square lattices to calculate $\Omega(E)$. Bhanot's method is 
far more complex than the $\mu TM$ and requires essentially
the same computer resources. Pearson's method is only applicable
to lattices with very few spins (e.g. 64) and Beale's approach
makes essential use of the exact solution for the Ising model
and so can not be used for other values of $Q$ or in three
dimensions. The $\mu TM$ is quite general and the algorithm
itself requires less than 100 lines of code. In addition,
it is straightforward to generalize the $\mu TM$ to count
states with fixed energy {\it and magnetization}, or 
any other function of the Potts variables.

\section{Results}

The partition function for the Potts model maps into itself
under the dual transformation
$$u\to{1\over u},\eqno{(7)}$$
where
$$u={y^{-1}-1 \over \sqrt{Q}}.\eqno{(8)}$$
In the complex $u$-plane a subset of the zeros of the partition
function tend to lie on a unit circle (which maps into itself
under (7)); however, cylindrical and periodic boundary conditions
are not self-dual, and this causes the zeros to move slightly off
the unit circle.
For this reason we have modified the $\mu TM$ for the self-dual
lattice introduced by Wu {\it et al.}\cite{wu}, so that
the zeros do indeed lie on the unit circle
and therefore are simply parameterized by the phase $\theta$.

In the complex $x$-plane the zeros of the partition function for 
the Ising model are guaranteed to lie on the unit circle
by the circle theorem of Lee and Yang\cite{ly}, irrespective
of the boundary conditions.

Given that the zeros are well parameterized by a single variable,
we can define the density of zeros for finite lattices as
$$g({1\over 2}(\theta_{k+1}-\theta_k))
={1\over N}{1\over\theta_{k+1}-\theta_k}.\eqno{(9)}$$  
In Fig.1 we show the density
of zeros in the complex-$x$ plane for the Ising model at 
$y=y_c$ and $y=0.5y_c$. Note that at the critical temperature
$g$ tends to zero as one approaches the real axis, but below the 
critical temperature it approaches a constant
$$2\pi g(0,y)=m_0(y),\eqno{(10)}$$
where $m_0(y)$ is the spontaneous magnetization.
We have applied finite-size scaling to the density of zeros
calculated in this way and
find excellent agreement with the exact solution for the 
magnetization except close to the critical point where crossover
complicates the FSS analysis. There is reason to hope that
a more sophisticated FSS analysis will improve these results
substantially.

Finally, in Fig.2 we show the density of zeros 
in the complex $y$-plane for the 3-state Potts model,
which is known to have a second-order transition
at the critical point.

\section{Conclusions}

The $\mu TM$ and its extensions offer a new way of obtaining
exact information about finite two dimensional lattices.
While the method is easily extended to three dimensions,
memory requirements limit its use to the 2-state model on
$4^2\times L$ lattices. However, Monte Carlo techniques
have been developed\cite{cp} which have no such limitations 
and show great promise in extending many of these results
to larger lattices. Preliminary studies indicate that 
while most of the Yang-Lee zeros are very sensitive
to the exact value of $\Omega(E)$, the edge singularity
and the next two nearest the critical point are not.

In addition, it is possible to generalize the $\mu TM$
to calculate the microcanonical distribution of any function
of the Potts variables, and in particular the order parameter
and correlation function.

\begin{figure}
\epsfxsize=12cm\epsfbox{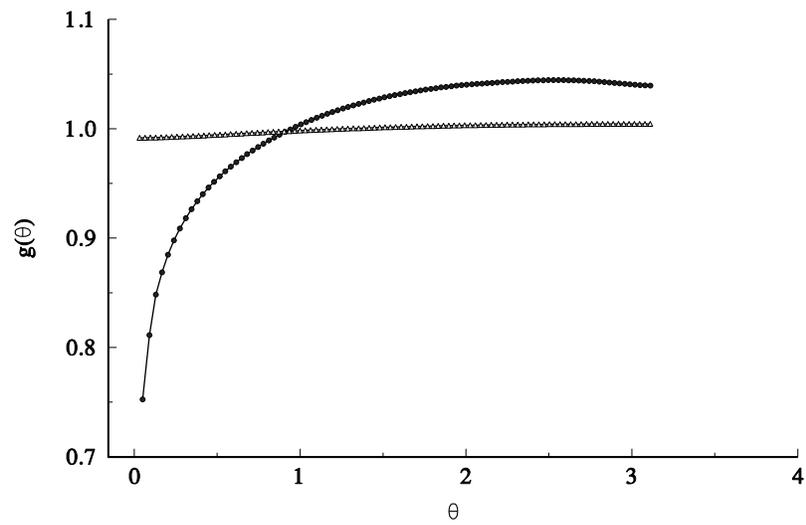}
\caption{Density of zeros in the complex $x$-plane for the Ising model
for $L=14$, cylindrical boundary conditions,
for $y=y_c$ (filled circles) and $y=0.5y_c$ (open triangles).}
\end{figure}

\begin{figure}
\epsfxsize=12cm\epsfbox{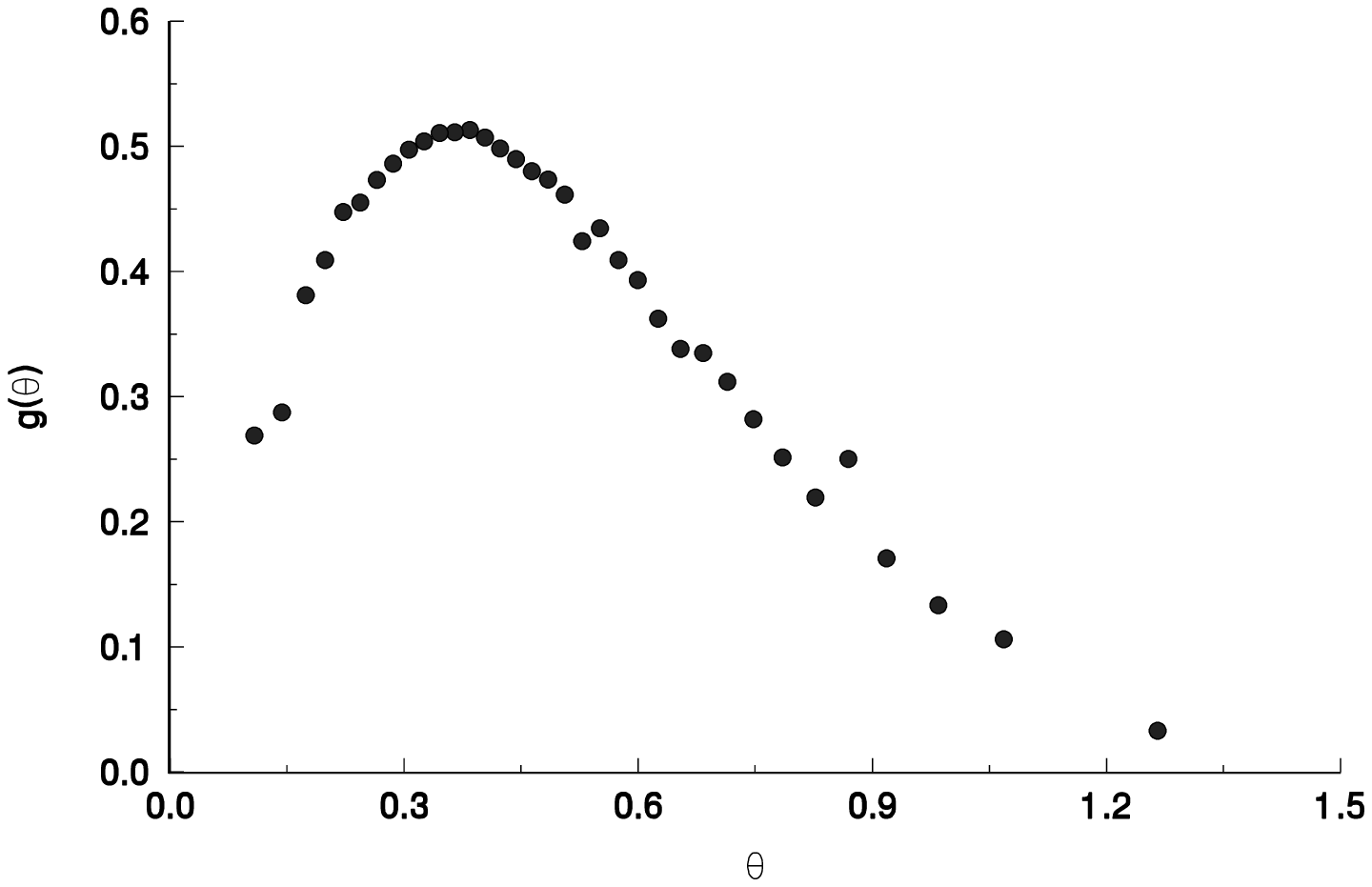}
\caption{Density of zeros 
in the complex $y$-plane for the 3-state Potts model
for $L=10$, self-dual boundary conditions.}
\end{figure}


\begin{thebibliography}{99}
\bibitem{yl} C. N. Yang and T. D. Lee, 
Phys. Rev. {\bf 87}, 404 (1952).
\bibitem{ly} T. D. Lee and C. N. Yang, 
Phys. Rev. {\bf 87}, 410 (1952).
\bibitem{ab} R. Abe,
Prog. Theor. Phys. {\bf 38}, 72 (1967).
\bibitem{su} M. Suzuki,
Prog. Theor. Phys. {\bf 38}, 1225 (1967).
\bibitem{ck} R. J. Creswick and S.-Y. Kim, 
Phys. Rev. E {\bf 56}, 2418 (1997).
\bibitem{cr} R. J. Creswick,
Phys. Rev. E {\bf 52}, R5735 (1995).
\bibitem{bi} K. Binder,
Physica {\bf 62}, 508 (1972).
\bibitem{bh} G. Bhanot,
J. Stat. Phys. {\bf 60}, 55 (1990).
\bibitem{pe} R. B. Pearson,
Phys. Rev. B {\bf 26}, 6285 (1982).
\bibitem{be} P. D. Beale,
Phys. Rev. Lett. {\bf 76}, 78 (1996).
\bibitem{wu} C.-N. Chen, C.-K. Hu, and F. Y. Wu,
Phys. Rev. Lett. {\bf 76}, 169 (1996).
\bibitem{cp} G. Bhanot, R. Salvador, S. Black, P. Carter, 
and R. Toral, Phys. Rev. Lett. {\bf 59}, 803 (1987);
K.-C. Lee, J. Phys. A {\bf 28} 4835 (1995);
C. M. Care, {\it ibid.} {\bf 29} L505 (1996);
A. Pavel'yev, Ph.D. thesis, University of South Carolina, 1997 (unpublished).
\end{thebibliography}
\end{document}